# On Computational Infraestruture Requirements to Smart and Autonomic Cities Framework

Romildo M. S. Bezerra, Flávia M. S. Nascimento
Grupo de Pesquisa em Sistemas Distribuídos, Otimização, Redes e Tempo Real
Federal Institute
Salvador, Bahia
(romildo,flaviamsn)@iba.edu.br

Joberto S. B. Martins, IEEE Senior Member
Núcleo de Pesquisa em Redes de Computadores
Salvador University
Salvador, Bahia
joberto@unifacs.br

*Abstract*— Smart cities are an actual trend being pursued by research that, fundamentally, tries to improve city´s management on behalf of a better human quality of live. This paper proposes a new autonomic complementary approach for smart cities management. It is argued that smart city management systems with autonomic characteristics will improve and facilitate management functionalities in general. A framework is also presented as use case considering specific application scenarios like smart-health, smart-grid, smart-environment and smart-streets.

*Keywords—autonomic systems; smart cities; management.*

## I. Introduction

Population growth in cities results in growing urbanization and economy evolution and this urban expansion is one of the biggest building boom humanity will ever undertake [7]. In 1800, only 3 % of the world population lived in cities. It is estimated that this number will reach 65% in 2040 (Figure 01). With the global population increase and migration phenomena of people to the cities in search more resources, it is expected that around 30 cities will have more than 10 million inhabitants by 2040 As another example of the huge numbers involved, the largest cities consume 85% of the world's energy and produce 80% of its waste.

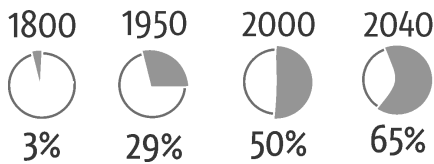

Fig. 1. Population growth in large cities

At the actual "technological moment", we can live in Smart Cities full of sensors and machines that talk to other machines on their own helping us to improve our lifestyle. As a result, people and objects jump into the "new" Internet (Internet of Things) adding new layers of data and complexity [l]. Thus, a Smart City should potentially enable citizens to engage with various new services on offer, public or private, in a way best suited to all [2].

In [2], it is presented five important aspects to improve "smartness" in cities:

- A robust and scalable digital infrastructure, combined with a secure but open access approach to public findable, useable and sharable data, which enables citizens to access the information they need, when they need it;
- The recognition that service in general could be improved by being "citizen centric";
- An smarter physical infrastructure (devices, networks and systems) is required in order to enable service providers to manipulate massive volumes of data and manage service delivery;
- An openness to learn from others and experiment with new approaches and new business models;
- Transparency of performance.

Thus, in this paper a "smart city" is considered as the one that uses digital technologies to improve its performance management and well-being, reduces costs and resource consumption and engages more efficiently and effectively with its citizens.

To meet such requirements described and allow at the same time an improvement in the quality of life, thousands of sensors and smart devices will be distributed collecting raw data continuously. If each sensor collects a few tens of kilobytes and send them every minute, at the end of a day, a few terabytes of raw data will be stored (without relationship with each other and without history matching) probably in different databases.

In this context, one possible approach is to develop a "framework" that enables the human activity optimization to minimize the high complexity of smart cities management. In this same scenario, it is considered that "autonomic computing" can contribute and collaborate in the management of the smart cities complexity.

This paper is organized as follows: In Section II, we present basic concepts of autonomic computing; The Section III presents a proposal for implementing autonomy in smart cities context, indicating challenges and requirements. In Section IV, the final remarks are presented.

## II. AUTONOMIC COMPUTING: AN BRIEF OVERVIEW

Autonomic Computing is an inevitable evolution of the information technology infrastructure management [3]. This evolution was necessary because the complexity of computing environments has increased due to greater sophistication of services offered, the demand for quality and productivity, the data growing and the heterogeneity of device, technologies and platforms.

Actual complex infrastructures are difficult to management, making manager's activities simultaneously costly, complex and error prone. The complexity is seen presently as the most important challenge to be addressed by the autonomic management systems [4].

In complex systems, the human intervention may be a fault insertion point as far as the management is concerned [5]. In this regard, an autonomic management system can be perceived as a less human-dependent entity that runs most administrative routines and operational tasks [4]. However, it is important to point out that autonomic computing does not focus on the elimination of human intervention [26], but in a high-level participation to set goals and business rules to be followed by such systems. This way, you can see autonomic computing as a system that has the ability to manage any system in accordance with the objectives set by the administrator. [6]

In fact, the essence of autonomic systems is Self-Management [6], which aims to make the managed environment able to perceive, analyze its current conditions and have the ability to reconfigure its components and devices proactively.

For Self-Management capabilities to be achieved over any managed infrastructure, four key requirements must be met [6]:

- Self-Configuration: Consists of the system's ability of adapt to managed environment conditions, predictable or not, dynamically adjusting its configuration (on-the-fly). Self-Configuration is not limited to the system ability to configure each device separately, but rather to provide the ability to adjust the set of devices dynamically and globally for the well-being of the environment managed as a whole.

- Self-Optimization: Autonomic systems should be able to seek the improvement of its operation. This could be realized by identifying new opportunities (states) that allows the execution of same operation with lower cost, better performance or using fewer resources. For this, they should implement a proactive search in order to identify, verify and make configuration changes that are able to maximize the use of resources without human intervention.

- Self-Healing: The Self-healing capability can prevent and recover from a failure, searching, diagnosing and fixing points that can cause interruption in the services offered. For this, the system should be able to isolate a faulty device or component in order to minimize the impact on services, continuously maximizing the availability and reliability of the managed environment.

- Self-Protection: Corresponds to the capacity to detect, identify and protect against attacks and treats. Faults and problem foreseen is a key feature needed in self-protection. This is typically done by correlating data and/or analyzing previous system states. Issues and aspects evaluated in order to foreseen attacks include unauthorized access, virus detection, DoS attacks and general faults among many others. Self-Protection feature also includes the capability to recognize overload situations that could eventually compromise the integrity of the managed system.

- Self-Healing: Is the ability to detect, identify and repair/fix (healing) against attacks and/or undesirable situations.

In architectural terms, an autonomic system can be divided into four phases (Figure 02) [8]:

- Monitoring: Allows the autonomic manager to collect, aggregate, filter and report details such as metrics or topologies, from the device(s)/ system managed.

- Analysis: This step checks the current state of the system based on the monitoring data. In general, this phase requires the use of complex systems for evaluation of several possible system conditions.

- Planning: It consists in defining the set of actions required to achieve the high-level goals/objectives.

- Execution: This feature allows the autonomic manager to change the managed resource behavior.

Given this brief information, it can be seen that the smart cities management scenario fits the profile and perspectives of autonomic computing activities, by considering that complexity, dynamism and heterogeneity are characteristic belonging to both contexts. As such, we present in the following section an autonomic computing framework for smart city management through a prototype called SACI – Smart and Autonomic Cities – which has focused on the environment smart management, smart streets, smart health and smart grids.

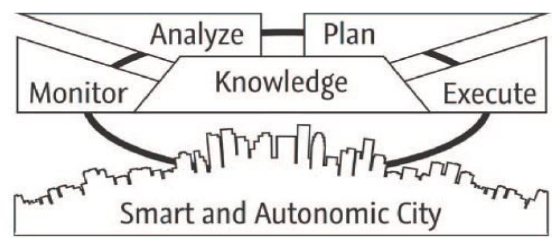

Fig 2 – Smart and Autonomic City Overview

## III. Applying Autonomy in Smart Cities Context

Autonomic computing principles can be used in the context of smart cities, considering that such environments are complex, unpredictable and large scale. The goal of applying the autonomic computing approach to smart cities context is to simplify the management process and reduce human intervention, seen by many as a point of failure [5].

Arising from this, one of the objectives of this article is to propose the implementation of autonomy in smart cities, emphasizing its requirements, challenges and computing infrastructure required.

The second objective is to present the Smart and Autonomic City Framework (SACi Project) as a case study.

### A. SACi Project: Smart and Autonomic City Framework

The SACi Project (Figure 3) is a high-level integration framework composed by four management systems with autonomic characteristics for different city´s smart-properties. The SACi framework integrates currently available and/or under-development autonomic systems supporting smart-health, smart-grid, smart-environment and smart-streets.

- Smart-Health: The health monitoring systems fundamentally focus on enabling more autonomic remote monitoring when physical presence is not possible or is difficult to implement. This suits the reality of many metropolitan cities with both remote areas and, in some cases, with serious social problems. As is well known in large urban centers, the logistics for supporting patients is complex and mostly compromised by chaotic urban traffic that frequently compromise the quality of care [9].

The patients monitoring system is called WHMS4 (Simple Wearable Health Monitoring, and Scalable Secure System) [9] [10]. It uses a set of sensors, microcontroller, speaker and other devices coupled to the clothing (wearable) in order to support treatment and patients monitoring.

The main WHMS4 objective is to provide health status information of elderly people via Internet, using wearable computing devices, customizable and low cost.

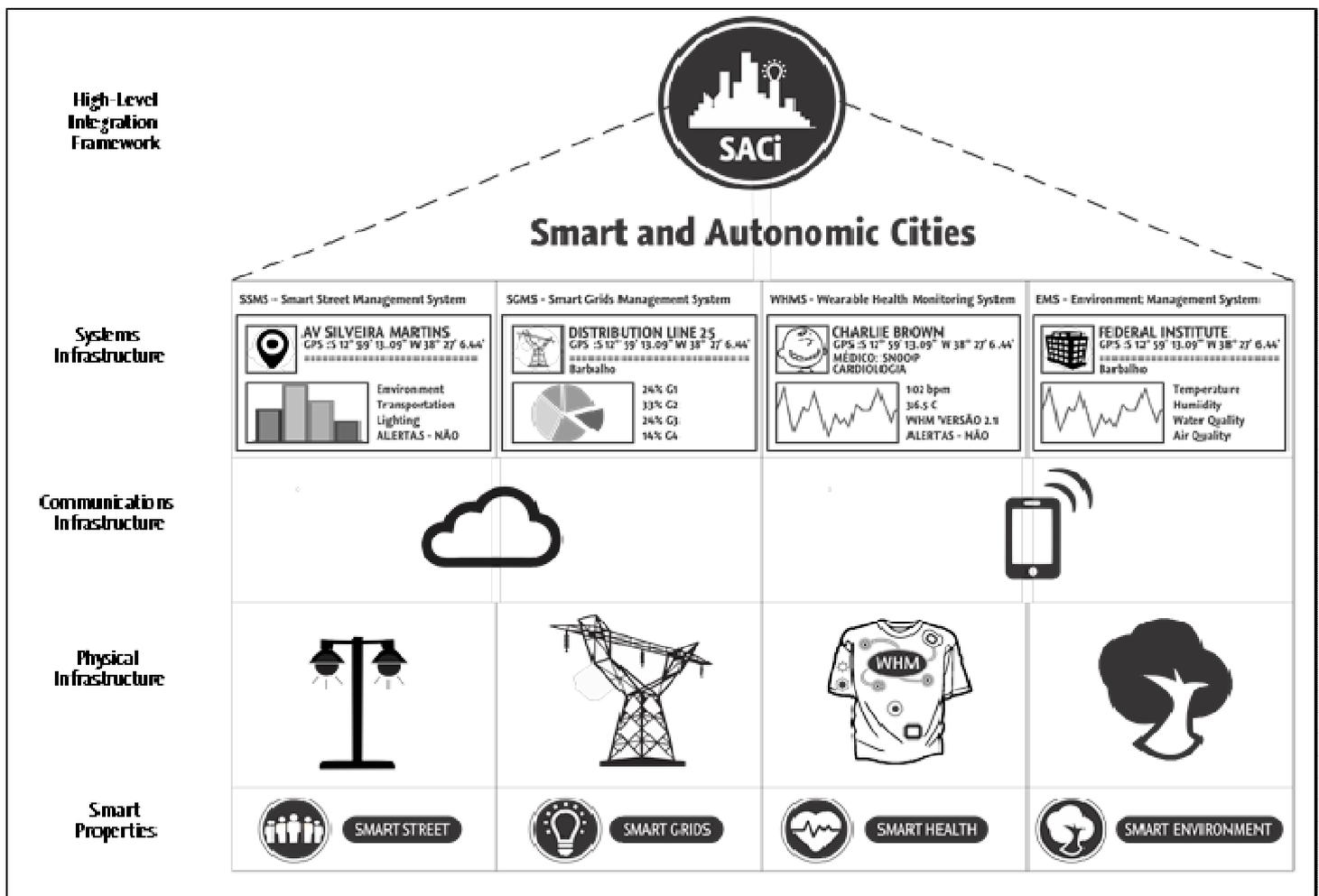

Fig. 3. Smart and Autonomic Cities Framework

- Smart-Grids: Smart-grid is a generic term meaning the application of computational intelligence and networking capabilities to dumb electricity distribution systems [24]. In general, you can apply intelligence at different grid levels (from the supply to the consumption of electricity). The specific focus of SGMS (Smart Grids Management Systems) [27] is to equalize the contribution of generators to the city distribution lines allowing the improvement of load balancing [25].

- Smart-Environment: The smart-environment systems, in general, focus on: (i) air quality[1]; (ii) water quality[2] in rivers, lakes and the sea; (iii) quality in urban spaces[3]. It is a traditional smart property and exists many projects and solutions. The EMS (Environment Management System) project specifically develops and proposes the use of simple Arduino devices (nano) and sensors in closed environment contexts.

- Smart-Streets: Smart-street system proposes and introduces a new concept for smart-environment by aggregating: (i) Social and Human Capital; (ii) Transport and Mobility information; (iii) Quality of live; (iv) Public Lighting Control. In order to support the provision of services, it is proposed the use the lampposts as smart device in urban spaces. The prototypes are equipped with Intel Galileo 2, speakers, microphone, LCD screen and sensors. The main objective is to offer information to citizens as: transportation, mobility, lighting, security, tourism, etc.

B. *Autonomic Challenges in Smart Cities Context*

Smart cities present a series of challenging requirements considered next.

- Heterogeneous functionality management [11]: An issue and requirement existent in the implementation of autonomic principles for smart cities management is the availability of multiple devices with distinct resources (capabilities). Due to the heterogeneity of models and consequent failure to meet standards, smart cities devices end up with different representation of their resources and, typically, have different monitoring facilities. Therefore, it is necessary to abstract specific device features in order to facilitate a standard way to reconfigure and manage devices. This abstraction facility will facilitate self-configuration and will allow non-autonomic features to be administered by autonomic systems.

- Reliability [15]: In autonomic systems context, the system reliability can be measured by the actions correctness taken during the run. A reliable autonomic system is one that works without interruption, as expected or promised (based on business rules), providing an adequate service whenever possible. Lack of reliability can affect the on-the-fly smart cities system solutions.

- Scalability [8] [14]: Autonomic systems are used for managing complex systems, regardless of the number of managed devices, which may vary from tens to thousands of elements. One actual approach used to deal with the scalability issue is to consider "domains" independently of their complexity. As such, an inappropriate domain choice can affect the quality of the solution for the autonomic system. In this same context, the "quality of the solution" can be understood as the measurement of SLAs met.

- Robustness [28]: A robust system can be defined as a tolerant system to undesirable state, i.e. a robust fair persistent operation, even with continual disruptions, failures, attacks or degrading behavior (due to the action of users and applications). This point is of fundamental importance to the high availability of a smart city management system.

- Adaptability [12] [13]: Adaptability is the ability of the system to adapt in response to faults change in user´s (application) requirements, business rules, and/or environmental conditions such as faults in devices distributed in a city. Adaptability in smart cities is also related to the need for adaptation of a management system will different cities, either by size, complexity, characteristics or economy. In this context, the more adaptable is the system, the greater its complexity.

- Application of learning and reasoning techniques to support intelligent interaction [12]: Systems must be able to understand the current state of smart-property being analyzed by the autonomic management system. For example, device´s static data can be collected and analyzed to determine whether a problem has been or can be optimized. Note that the management's current data do not report to the administrator, for example, why a traffic jam and the same congestion may not have the same cause. Thus, this information must be inferred from the correlation data (reasoning) and, if possible, a study of previous cases. If the system is also able to save information for future reference (learning), such a system will be heading towards the state of the art in the context of autonomy for smart cities.

In non-autonomic management, administration of a smart city is performed typically by a group of IT professionals that models the system according to the needs of the administrator group. Access to the system is decentralized according with the specialty to be managed. However, this reactive profile and decentralized intelligence typically does not meet the expectations of the design of a smart city and their complexity.

In simple terms, the high-level model of an autonomic system can be a black box that receives the environmental data (such as sensors) and imports the business rules specified by managers, generating or optimizing a solution to the problem occurred.

---

[1] Temperature, Humidity, Carbon Monoxide (CO), Carbon Dioxide ($CO_2$), Oxygen ($O_2$), Methane ($CH_4$), etc.
[2] Hydrogen ($H_2$), Ammonia ($NH_3$), etc.
[3] Dust concentration, Ultrasounds, Noise, Luminosity, etc.

To define the operation of an autonomic architecture for smart cities it is necesary specify its requirements and desirable characteristics. The basic requirements set for an architecture with autonomic features for smart cities are:

- Scalability: In smart city management there is a tendency to have a high number of devices and communications equipment. Thus, the autonomic architecture should be able to act in different cities cardinalities and allow significant expansion that can affect a larger volume of data managed.

- Modularity: It is an important requirement in the development of an autonomic architecture in general and should also be associated with the smart cites context. The smart city autonomic framework must be able to support different algorithms, protocols and standards, without interfering operation. Moreover, the overall structure of the solution should be such that not adopt a monolithic structure oriented only to the displayed application focus. In [23] it is presented a modular framework based in trace analysis and mining that can reveal exact and inherent information or knowledge about the city and its people to the benefit of many applications such as transportation, health or security.

- Self-Stabilization: It is a concept that has emerged as a promising paradigm for the design, control and maintenance of distributed systems tolerant to failure, because it allows systems to automatically recover upon the occurrence of failures [16]. Its basic idea is to maintain the expected behavior of an autonomic system, regardless of the state in which it finds itself, even though a failure may put it in an arbitrary state (not expected). In this case, the system may possibly (if resources are available) resume its expected behavior. This concept was introduced by Dijkstra [17], who defined self-stabilization with a characteristic of systems that reach a legitimate state[4], regardless of its initial state in a finite number of steps.

- Real-Time Requirements: A real-time system is a computational system in which both timing and logical requirements must be respected. For this, it is necessary to offer correctness and timeless [29]. In autonomic cities context, a real-time implementation needs important network requirements (bandwidth and delay), hardware (battery) and traditional scheduling algorithms (RM[5] and EDF[6]) in order to guarantee timeliness.

- Learning and Reasoning: As previously mentioned, autonomic systems need to have a search engine solution and, if possible, to learn the solutions previously performed. Several architectural proposals have been used in other application areas of autonomic computing as:

    o Inspired Living Systems Design [18]
    o Bio-Inspired Survivability [19]
    o Collective Behavior [20]
    o Policy-Based Design [21] or
    o Knowledge Plane Design [22].

Regarding learning technique used in autonomic systems, according to [13], during the phase of analysis and planning various techniques can be used such as inference, game theory, economic models, fuzzy logic, etc. The event correlation can also be a technique used for integrated management in autonomic systems because it uses "past experience" to find new solutions by techniques such RBR (Rule-Based Reasoning) and CBR (Case-Based Reasoning). In the context of Autonomic Cities, the CBR can be used in approaches to solve event-based problems as it has the ability to automatically adapt to find new solutions based on past cases, as well as human thinking.

Thus, a viable proposal for the implementation of autonomy in smart cities corresponds to create a high-level framework, integrated with other smart systems management properties, which is responsible for management decision making in accordance with the requirements placed by managers. The smart-properties management systems are responsible for the application of predefined rules, while the definition of such rules is supported by the framework. (Figure 4).

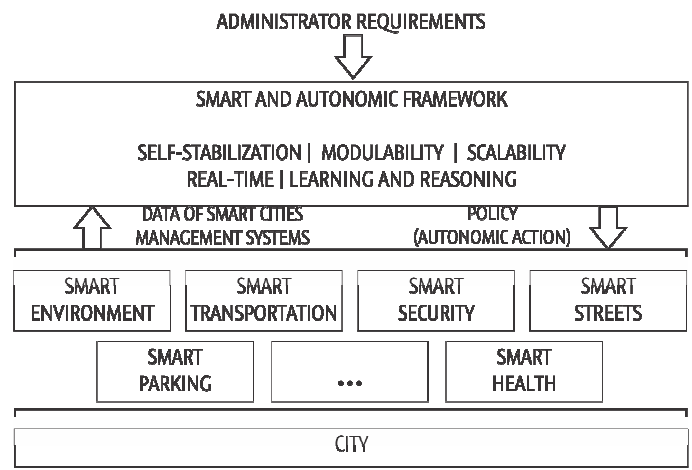

Fig. 4. Modeling of smart cities framework with autonomic features.

## IV. Final Considerations

Smart city systems (smart transportation systems, smart grid for cities, environment smart systems, smart water systems, smart-health systems, other) with more autonomic characteristics is an expected capability and actual trend in Future Internet (FI) evolutionary path scenario for cities.

---

[4] A state is considered legitimate when its operation meets the expected requirements.
[5] Rate Monotonic
[6] Earliest Deadline First

Considering this scenario, the goal pursued is towards the "Smart and Autonomic Cities".

This paper evaluates the requirements imposed to the computational infrastructure beneath smart city systems considering the introduction of new autonomic management characteristics to them.

A framework (SACI – Smart and Autonomic Cities) is presented integrating distinct systems as a use case adopted in order to map and guide the discussion about the computational infrastructure autonomic requirements. It is argued that the introduction of autonomic management characteristics for city systems must consider basic autonomic capabilities such as adaptability, self-stabilization, modularity, real-time and learning and reasoning in a context with inherent complexity, high volume of data and heterogeneous devices involved.